\documentclass[sigconf]{acmart}

\usepackage{xspace}
\usepackage{amssymb,amsmath} 
\usepackage{algorithmic}
\usepackage{algorithm}
\usepackage{balance}
\usepackage{enumitem}
\usepackage{tikz}

\AtBeginDocument{%
  }

\copyrightyear{2025}
\acmYear{2025}
\setcopyright{acmlicensed}\acmConference[CIKM '25]{Proceedings of the 34th ACM International Conference on Information and Knowledge Management}{November 10--14, 2025}{Seoul, Republic of Korea}
\acmBooktitle{Proceedings of the 34th ACM International Conference on Information and Knowledge Management (CIKM '25), November 10--14, 2025, Seoul, Republic of Korea}
\acmDOI{10.1145/3746252.3761194}
\acmISBN{979-8-4007-2040-6/2025/11}
\newcommand{\sysname}{\textit{PP-STAT}\xspace}
\newcommand{\hstat}{\textit{HEaaN-STAT}\xspace}
\newcommand{\panda}{\textit{Pivot-Tangent}\xspace}

\newcommand{\invSqrt}{\textit{CryptoInvSqrt}\xspace}

\begin{document}

\title{\sysname: An Efficient Privacy-Preserving Statistical Analysis Framework using Homomorphic Encryption}

\author{Hyunmin Choi} 
\orcid{0009-0002-0486-9582} 
\affiliation{%
\institution{NAVER Cloud \& Sungkyunkwan University}
\city{Seongnam} 
\country{South Korea}} 
\email{hyunmin.choi@g.skku.edu}

\renewcommand{\shortauthors}{Hyunmin Choi}

\begin{abstract}
With the widespread adoption of cloud computing, the need for outsourcing statistical analysis to third-party platforms is growing rapidly. However, handling sensitive data such as medical records and financial information in cloud environments raises serious privacy concerns. 
In this paper, we present \sysname{}, a novel and efficient Homomorphic Encryption (HE)-based framework for privacy-preserving statistical analysis. HE enables computations to be performed directly on encrypted data without revealing the underlying plaintext. \sysname{} supports advanced statistical measures, including Z-score normalization, skewness, kurtosis, coefficient of variation, and Pearson correlation coefficient, all computed securely over encrypted data.
To improve efficiency, \sysname{} introduces two key optimizations: (1) a Chebyshev-based approximation strategy for initializing inverse square root operations, and (2) a pre-normalization scaling technique that reduces multiplicative depth by folding constant scaling factors into mean and variance computations. These techniques significantly lower computational overhead and minimize the number of expensive bootstrapping procedures.
Our evaluation on real-world datasets demonstrates that \sysname{} achieves high numerical accuracy, with mean relative error (MRE) below $2.4 \times 10^{-4}$. Notably, the encrypted Pearson correlation coefficient between the \texttt{smoker} attribute and \texttt{charges} reaches 0.7873, with an MRE of $2.86 \times 10^{-4}$. These results confirm the practical utility of \sysname{} for secure and precise statistical analysis in privacy-sensitive domains.

\end{abstract}

\ccsdesc[500]{Security and privacy~Domain-specific security and privacy architectures}
\ccsdesc[500]{Mathematics of computing~Probability and statistics}

\keywords{Homomorphic Encryption, CKKS scheme, Statistical analysis, Privacy-preserving computation}


\bibliographystyle{ACM-Reference-Format}
\balance
\maketitle
\section{Introduction}
\label{sec:introduction}
Statistical analysis services are increasingly deployed across domains such as finance, education, and healthcare.  
These services rely on core statistical measures—such as Z-score normalization, coefficient of variation, skewness, kurtosis, and Pearson correlation coefficient—to extract meaningful patterns from data~\cite{campisi2023assessing,akter2018investigation,pamuk2022role,malehi2015statistical,de2016comparing}.

To scale these services, cloud computing has become a de facto solution. However, outsourcing sensitive data to public cloud platforms raises serious privacy concerns, including the risk of unauthorized access or data leakage. To mitigate such threats, homomorphic encryption (HE) enables computation directly over encrypted data without decryption, supporting a variety of privacy-preserving machine learning and statistical workloads~\cite{dathathri2019chet,ishiyama2020highly,choi2024unihenn,lee2023heaan}.

Among existing HE schemes, the Cheon-Kim-Kim-Song (CKKS) scheme~\cite{cheon2017homomorphic} is particularly suited for real-valued computations, making it a practical foundation for encrypted analytics. However, despite its flexibility, CKKS poses significant implementation challenges: it only supports polynomial operations (addition, multiplication, and rotation), requiring complex polynomial approximations for non-linear functions. As a result, managing ciphertext noise and computational depth becomes a critical bottleneck in large-scale HE-based statistical analysis.

Despite growing interest, existing HE-based frameworks for data analysis are often closed-source or limited in scope, making practical adoption difficult. To address these limitations, we present \sysname, an efficient and scalable open-source framework for encrypted statistical analysis. \sysname implements five foundational statistical operations—Z-score normalization, coefficient of variation, skewness, kurtosis, and Pearson correlation coefficient—executed securely under the CKKS scheme.
Many statistical measures in \sysname rely on the inverse square root operation, which is non-trivial to implement efficiently under HE.  
Existing methods such as Goldschmidt or Newton’s method~\cite{cetin2015arithmetic} have been adapted for HE, but typically assume fixed initial guesses. 
For example, Lee et al.~\cite{lee2023heaan} use $y_0 = 1$ as the initial value in Newton’s method, which leads to slow convergence and reduced accuracy for small $x$. Panda et al.~\cite{panda2022polynomial} improved initialization via a homomorphic \textit{sign} function (Pivot-Tangent method), but the approach requires six bootstrapping operations. 

To overcome these limitations, we propose \invSqrt, a Chebyshev-based initialization method for Newton’s iteration under HE. It achieves faster convergence while significantly reducing bootstrapping overhead. Specifically, \invSqrt requires only two bootstrapping operations, which is 2.5× and 3× fewer than those used in Lee et al.~\cite{lee2023heaan} and Panda et al.~\cite{panda2022polynomial}, respectively. On the interval $[0.001, 100]$, \invSqrt achieves a mean relative error (MRE) of $5.08 \times 10^{-5}$, outperforming Pivot-Tangent (MRE: $3.73 \times 10^{-4}$) while using less than half its bootstrapping operations. 
We further evaluate \sysname on two real-world datasets: \textit{Adult} and \textit{Insurance}. When computing the encrypted Pearson correlation coefficient between the \texttt{smoker} and \texttt{charges} attributes in the \textit{Insurance} dataset, \sysname produces a value of 0.7873, with a mean relative error of $2.86 \times 10^{-4}$. This result demonstrates that accurate and meaningful statistical analysis can be conducted directly over encrypted data without revealing sensitive information.

Our contributions are as follows:
\begin{itemize}
    \item We present \sysname, an efficient and scalable HE-based framework that supports five advanced statistical operations: Z-score normalization, kurtosis, skewness, coefficient of variation, and Pearson correlation coefficient—over encrypted data.
    \item We propose \invSqrt, a Chebyshev-based initialization for Newton’s method that reduces bootstrapping to two rounds with higher accuracy and faster runtime than prior work. We also introduce pre-normalization scaling, enabling higher-degree Chebyshev polynomials within fixed multiplicative depth to improve accuracy without extra cost.
    \item We empirically demonstrate that \sysname achieves high numerical accuracy on large-scale datasets. For example, Z-score normalization over one million records in the domain $[0, 100]$ yields an MRE of $4.18 \times 10^{-5}$.
    \item We validate the practical utility of \sysname through experiments on two real-world datasets, \textit{Adult} and \textit{Insurance}. In the \textit{Insurance} dataset, it identifies a strong correlation (Pearson correlation coefficient $=0.7873$) between \texttt{smoker} and \texttt{charges}, with an MRE of $2.86 \times 10^{-4}$.
    \item The full implementation of \sysname is publicly available at \url{https://github.com/hm-choi/pp-stat}. All experiments are fully reproducible and serve as a resource for further research and deployment.
\end{itemize}

\section{Related Work}
\label{sec:related_work}

\subsection{Inverse Square Root over HE}
Cheon et al.~\cite{cheon2019numerical} proposed HE-based methods for computing inverse and square root values, using Goldschmidt’s division algorithm~\cite{goldschmidt1964applications} and Wilkes’s algorithm~\cite{wilkes1958preparation}, respectively. These approaches rely solely on addition and multiplication, but do not provide a unified method for inverse $n$-th root computation, which is frequently required in statistical analysis.
Lee et al.~\cite{lee2023heaan} introduced HEaaN-STAT, a framework supporting various statistical operations under HE. For inverse $n$-th root, it applies Newton’s iteration with a fixed initial value, but does not optimize the initialization for convergence or depth reduction. Panda et al.~\cite{panda2022polynomial} proposed the Pivot-Tangent method, which estimates a better initial point for Newton’s iteration by evaluating a sign function. While this improves convergence, the sign function is computationally expensive under HE due to its non-linear nature.
Qu and Xu~\cite{qu2023improvements} identified the instability of minimax polynomials for Newton initialization, and instead proposed rational and Taylor expansion-based initializations of $1/\sqrt{x}$. These methods alleviate convergence issues but increase circuit depth. In contrast, our approach leverages Chebyshev approximation to generate accurate initial guesses, achieving stable convergence with fewer Newton iterations and significantly reduced bootstrapping overhead.

\subsection{Statistical Analysis over HE}
Several frameworks have been developed for encrypted machine learning. HElayers~\cite{helayers} supports neural networks and decision trees using a tensor-based abstraction. TenSEAL~\cite{tenseal2021} enables encrypted tensor computations for logistic regression and convolutional networks. UniHENN~\cite{choi2024unihenn} proposes an HE-friendly CNN inference architecture with approximation and batching strategies.
While effective for ML workloads, these systems provide limited support for general-purpose statistical analysis. In contrast, \sysname focuses on foundational measures such as Z-score normalization, skewness, and kurtosis, and incorporates Chebyshev-based inverse square root approximation—an optimization not available in existing frameworks.

\section{Background}
\label{sec:background}

\subsection{Homomorphic Encryption (HE)}
\label{subsec:homomorphic_encryption}

Homomorphic Encryption (HE) is an encryption scheme that allows computation on encrypted data without decryption. Following Gentry's blueprint of HE in 2009~\cite{gentry2009fully}, there have been many studies about HE. The Cheon-Kim-Kim-Song (CKKS) scheme~\cite{cheon2017homomorphic} is the fourth generation of HE that supports the encryption of a vector of real or complex numbers with predefined size. 
The CKKS scheme supports approximate arithmetic operations on real or complex vectors, including addition (Add), multiplication (Mul), and rotation (Rot), where Rot refers to cyclic slot-wise permutation. While rotation (Rot) is supported by CKKS and used implicitly during ciphertext evaluation (e.g., in bootstrapping), it is not discussed explicitly in this paper.
The allowed vector size is called the \textit{number of slots}. The allowed maximum number of multiplication of CKKS is called \textit{depth}, which is predefined in the key generation. If the number of multiplications exceeds the \textit{depth}, then the decryption result is not guaranteed. The remaining allowed number of multiplication in a ciphertext is called \textit{level} and denoted as $L$. The scale factor $\Delta$ guarantees the precision of the ciphertext. 
Add (C) and Mul (C) denote element-wise operations between two ciphertexts,  e.g., $Add(C(\mathbf{v_1}), C(\mathbf{v_2})) = C(\mathbf{v_1} \oplus \mathbf{v_2})$. Add (P) and Mul (P) operate between a ciphertext and a plaintext vector (or a constant).

\subsection{Bootstrapping in the CKKS Scheme}
\label{subsec:bootstrapping}
If a ciphertext's level reaches zero, further multiplication operations cannot be performed. To overcome this limitation, Gentry~\cite{gentry2009fully} proposed \emph{bootstrapping}, a technique that refreshes a ciphertext's level back to its initial ciphertext. Following Cheon~\cite{cheon2018bootstrapping}, numerous studies have been conducted to optimize the bootstrapping process in the CKKS scheme~\cite{chen2019improved,bae2022meta,bae2024bootstrapping}. 

However, bootstrapping in the CKKS scheme remains computationally expensive and time-consuming, as it involves encrypted Fourier transform operations. 

Table~\ref{table:op_comparision} summarizes the runtime (in milliseconds) of addition, multiplication, and bootstrapping operations in the Lattigo~\cite{lattigo} library. We set the ring degree to $N = 2^{16}$ with a scale factor of $\Delta = 2^{40}$. The total modulus size $\log_{2}(PQ)$ is set to 1443 bits, and the number of slots is $32{,}768$.

Each value represents the mean runtime over ten trials, with the standard deviation shown in parentheses.

\begin{table}[!h]
\caption{Runtime comparison of HE operations in Lattigo (depth = 11). All values are in milliseconds, averaged over ten trials.}
\centering
\small
\renewcommand{\arraystretch}{1.2}
\begin{tabular}{lccc}
\toprule
\textbf{Operation} & \textbf{Addition} & \textbf{Multiplication} & \textbf{Bootstrapping} \\
\midrule
Runtime (ms) & 6.41 (1.27) & 196.35 (6.69) & 43478.77 (302.74) \\
\bottomrule
\end{tabular}
\label{table:op_comparision}
\end{table}

As shown in Table~\ref{table:op_comparision}, bootstrapping requires more than 221.44$\times$ the runtime of multiplication.

\subsection{Newton's Method}
\label{subsec:newton_method}
The Newton's Method (Newton-Raphson Method~\cite{akram2015newton}) is an incremental algorithm to find a root $\alpha$ of a given function $f$. It starts with an initial value $x_{0}$, calculate the updating equation 
$x_{n+1} = x_{n} - f(x_{n})/f^{\prime}(x_{n})$ until the error between the estimated value $x_{n+1}$ and the real value $\alpha$. In this section, we only consider Newton's method to find an inverse square root. It is described in Algorithm~\ref{alg:newtion_iteration}).

\begin{algorithm}[!ht]
\caption{\textit{Newton's method for the inverse square root operation}}
\label{alg:newtion_iteration}
\begin{algorithmic}[1]
\STATE {\bfseries Input:} An initial point $x_{0}$, $y_{0}$ is an initial approximation at $x_{0}$, $d$ is the number of iterations
\STATE {\bfseries Output:} An approximation $y_{d}$.
\FOR{$i=1$ {\bfseries to} $d$}
    \STATE $y_{i} \leftarrow 0.5 \cdot y_{i-1} \cdot (3 - x_{0} \cdot y_{i-1}^{2})$
\ENDFOR
\STATE \textbf{return} $y_{d}$
\end{algorithmic}
\end{algorithm}

Newton's method converges very fast when the initial value $x_{0}$ is close to the real value $\alpha$, but if $x_{0}$ is far from $\alpha$, then it may diverge. Thus, it is important to find a good initial value $x_{0}$ for finding the suitable estimated result of the Newton's method.

\subsection{Polynomial Approximation}
\label{subsec:chebyshev_approximation}

The CKKS scheme allows arithmetic operations such as addition and multiplication, but only supports polynomial operations. Non-polynomial functions, such as square root or absolute value, cannot be directly supported in CKKS. Many previous approaches about implementing non-polynomial type of function use the approximation methods to convert it to a polynomial function. For instance, the approximated polynomial of the sign function can be obtained using the minimax approximation methods~\cite{cheon2020efficient,lee2021minimax,lee2022optimization}, and sine function used in the modulus evaluation in the bootstrapping obtained by Chebyshev approximation~\cite{chen2019improved,han2020better}. 

\subsection{Approximation Algorithm for Inverse Square Root}
To find an inverse square root of a given encrypted value, authors in \cite{panda2021principal} obtained a square root using Newton's method first with Goldschmidt’s algorithm to get the inverse of it. However, for fast convergence, Goldschmidt’s algorithm requires a suitable initial point of $1/\sqrt{x}$. 

Panda et al~\cite{panda2022polynomial} suggest a novel method \panda, a 2-line approximation to find a polynomial of an approximation of inverse square root using the pivot point and sign function. It overcomes the challenge of finding a good initial point for faster Newton's iteration, they split a domain $[a, b]$ as two intervals $[a, P]$ and $[P, b]$ as a pivot point $P$, and combine them using a function $\beta(x)=step(P/(b-a), x/(b-a))$ where $step(x) = (1+sign(x))/2$. The initial point $h(x)$ of $x$ using the Newton's method can be derived as follows:\\
$h(x) = (1-\beta(x))\cdot L_{1}(x) + \beta(x)\cdot L_{2}(x)$ where 
\begin{align*}
L_{1}(x) = -0.5 \cdot k_{2} \cdot x_{1}^{-1.5}\cdot x + 1.5 \cdot k_{2} / \sqrt{x_{1}} \\
L_{2}(x) = -0.5 \cdot k_{2} \cdot x_{2}^{-1.5}\cdot x + 1.5 \cdot k_{2} / \sqrt{x_{2}}
\end{align*}
In~\cite{panda2022polynomial} a method to find the pivot point $P$ and parameters $k_{1}, k_{2}, x_{1}, x_{2}$ and some examples of them are suggested. The convergence time of the algorithm is about half of them in~\cite{panda2021principal} using both Newton's method and Goldschmidt’s algorithm.

\section{Overview of \sysname}
\label{sec:overview}

\begin{figure}[!h]
\centerline{\includegraphics[width=0.9\columnwidth]{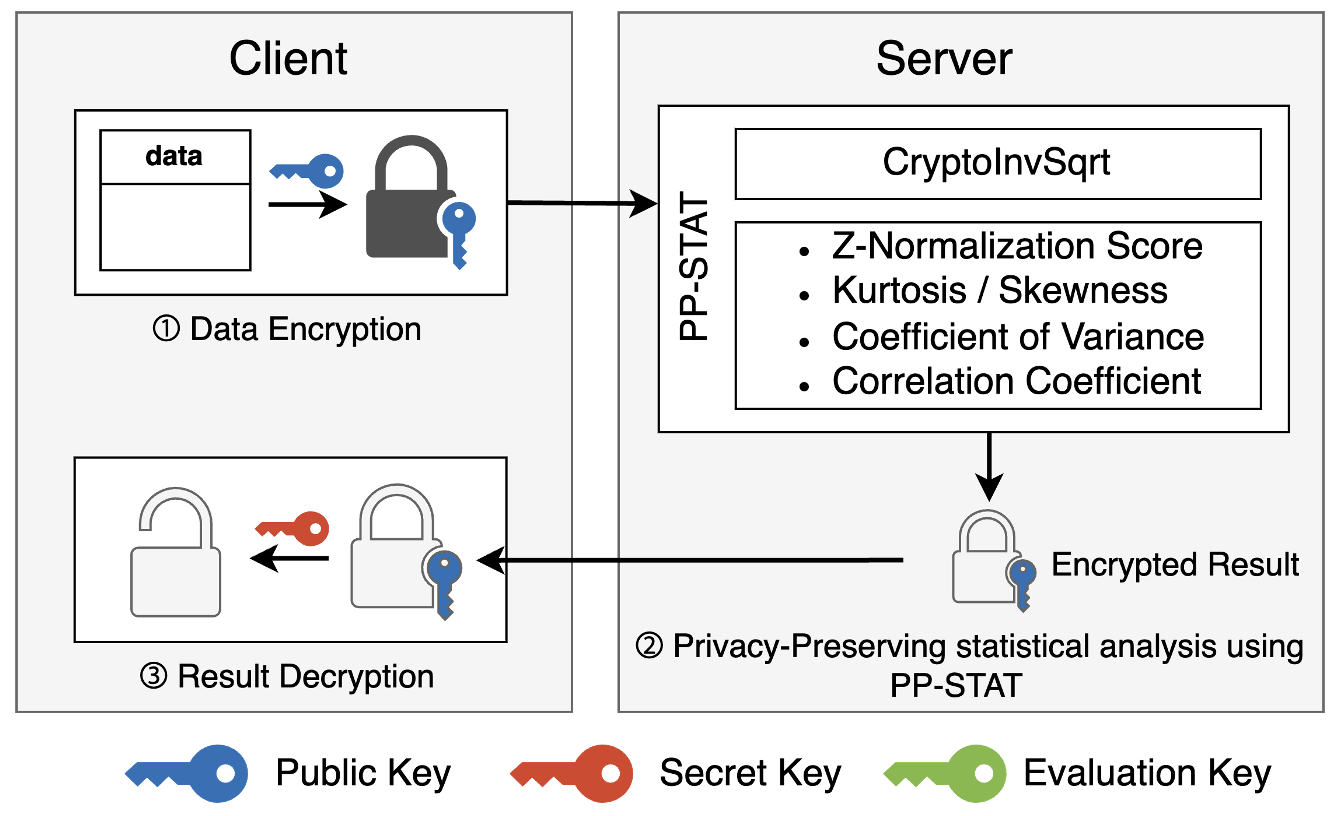}}
\caption{System overview of \sysname.}
\label{fig:overview}
\end{figure}

We present \sysname, a HE-based framework for privacy-preserving statistical analysis over encrypted client data. An overview of the system architecture is shown in Figure~\ref{fig:overview}.

\sysname supports advanced statistical operations including Z-score normalization, skewness, kurtosis, coefficient of variation, and Pearson correlation coefficient. These operations require inverse or inverse square root operations, which are not natively supported by the CKKS scheme. To address this, \sysname introduces an optimized Chebyshev polynomial approximation—\invSqrt for computing both $1/x$ and $1/\sqrt{x}$—and reuses them across all statistical functions to reduce depth and improve computational efficiency.

The system follows a standard client-server model:

\begin{enumerate}[label=\large\protect\textcircled{\small\arabic*}]
    \item The client encrypts sensitive input data using the public key and sends it to the server along with evaluation keys. The secret key is kept private and never leaves the client.
    \item The server performs statistical operations over the encrypted data using \sysname's supported operations.
    \item The client receives the encrypted results and decrypts them locally to obtain the final statistical outputs.
\end{enumerate}

Section~\ref{sec:construction} provides the technical details of the \invSqrt construction. Section~\ref{sec:stats} presents the algorithms used to compute the five supported statistical measures.

\section{HE-Based Inverse and Inverse Square Root Computation}
\label{sec:construction}
In this work, we define a unified function \invSqrt to represent the $n$-th inverse root operation of the form $x^{-1/n}$.  
This general form covers both inverse ($n=1$) and inverse square root ($n=2$), and is used throughout the paper for consistent construction and optimization under HE. The operation serves as a core computational primitives for several statistical functions in \sysname, including Z-score normalization, covariance, kurtosis, skewness, and Pearson correlation coefficient.

\subsection{HE implementation of the inverse square root}
\invSqrt is designed to estimate suitable initial values for Newton’s method in computing inverse $n$-th square roots under encryption. Notably, the inverse operation corresponds to the case where $n = 1$. Algorithm~\ref{alg:he_newton_method} describes a HE implementation of Newton’s method for approximating the inverse $n$-th square root of an encrypted input. The accuracy of this method critically depends on the initial guess $y_0$, which should closely approximate the true inverse $n$-th root of the encrypted input $x$. However, since $x$ is encrypted in the HE setting, selecting a suitable $y_0$ is challenging.

\begin{algorithm}[!ht]
\caption{\textit{HE-based Newton's method for inverse $n$-th square root}}
\label{alg:he_newton_method}
\begin{algorithmic}[1]
\STATE {\bfseries Input:} 
    \begin{itemize}
        \item $ct_{x}$: Ciphertext of $x$.
        \item $ct_{0}$: Ciphertext of $y_{0}$.
        \item $d$: Number of iterations
    \end{itemize}
\STATE {\bfseries Output:} 
    \begin{itemize} 
        \item $ct_{d}$: Approximation ciphertext.
    \end{itemize}
\IF{$n \geq 2$}
    \STATE $ct_{x} \leftarrow Mul(P)(ct_{x}, 1/n)$
\ENDIF
\FOR{$i=1$ {\bfseries to} $d$}
    \STATE $tmp_{a} \leftarrow Mul(P)(ct_{i-1}, (n+1)/n)$
    \STATE $tmp_{b} \leftarrow Mul(C)(ct_{x}, ct_{i-1})$
    \FOR{$j=1$  {\bfseries to} $n$}
        \STATE $tmp_{b} \leftarrow Mul(C)(tmp_{b}, ct_{i-1})$
    \ENDFOR
    \STATE $ct_{i} \leftarrow Sub(C)(tmp_{a}, tmp_{b})$
\ENDFOR
\STATE \textbf{return} $ct_{d}$
\end{algorithmic}
\end{algorithm}
Lee et al.~\cite{lee2023heaan} use a fixed constant for the initial value $y_0$ (e.g., $y_0 = 1$), which leads to slow convergence. Their method requires 25 and 21 Newton iterations for inverse $n$-th root with $n = 1$ and $n = 2$, respectively. This results in deep computation graphs and a total of five bootstrapping operations to maintain noise levels.

Panda et al.~\cite{panda2022polynomial} improve the initialization by applying a homomorphic \textit{sign} function, but their method still consumes six bootstrapping operations due to the additional depth required for polynomial evaluations and iterative refinement.

In contrast, \invSqrt approximates $y_0$ using a Chebyshev polynomial, which enables fast convergence in just 6 Newton iterations. Our method requires only two bootstrapping operations in total—less than half of Lee et al.’s and one-third of Panda et al.’s—while achieving higher accuracy and faster runtime. See Experiment 1 in Section~\ref{sec:experiment}.

\begin{algorithm}[!ht]
\caption{Initial value estimation for inverse $n$-th square root using Chebyshev approximation}
\label{alg:initial_value}
\begin{algorithmic}[1]
\STATE {\bfseries Input:} 
    \begin{itemize}
        \item $ct_{x}$: Ciphertext of $x$ (assumed in $[0, 1]$)
        \item $d$: Degree of Chebyshev approximation
        \item $n$: Root degree (e.g., 1 for Inv, 2 for InvSqrt)
    \end{itemize}
\STATE {\bfseries Output:} 
    \begin{itemize}
        \item $ct_{0}$: Initial approximation ciphertext
    \end{itemize}
\STATE $ct_{0} \leftarrow \text{ChebyshevApprox}(ct_{x}, d)$
\STATE $ct_{0} \leftarrow \text{Bootstrapping}(ct_{0})$
\IF{$n == 2$}
    \STATE $ct_{0} \leftarrow \text{Mul}(ct_{0}, ct_{0})$
\ENDIF
\STATE \textbf{return} $ct_{0}$
\end{algorithmic}
\end{algorithm}
In Algorithm~\ref{alg:initial_value}, a unified initial approximation is used for both inverse and inverse square root computations. Specifically, $\text{Inv}(x)$ can be computed by squaring $\text{InvSqrt}(x)$, eliminating the need for separate approximations.

\subsection{Domain Mapping for \invSqrt}
Since Chebyshev approximation is defined on the interval $[-1, 1]$, we set the function $F(x)$ as a shifted version of the inverse square root function to match this domain. The function is given as:

\begin{equation}
\label{eq:1}
F(x) =
\begin{cases} 
\frac{1}{\sqrt{x+1}} & \text{if } x > -1, \\
0 & \text{if } x \leq -1.
\end{cases}
\end{equation}

This definition ensures that $F(x)$ is continuous on $[-1, 1]$ and can be approximated using Chebyshev polynomials. The shift by $+1$ maps the original domain $[0, 2]$ of the inverse square root to the standard Chebyshev domain. A polynomial \( p_1(x) \) of fixed degree can be constructed to approximate \( F(x) \) over the domain \([-1, 1]\) using Chebyshev approximation. This transformation shifts the Chebyshev approximation back to the original inverse square root domain. As a result, $\mathrm{AppInvSqrt}(x)$ closely approximates $1/\sqrt{x}$ over $x \in [0, 2]$, while remaining Chebyshev-compatible in the computational pipeline.

\section{Statistical Operations with \invSqrt}
\label{sec:stats}
\sysname supports five advanced statistical operations: Z-score normalization, kurtosis, skewness, Pearson correlation coefficient, and coefficient of variation, as introduced in Section~\ref{sec:introduction}. All these measures fundamentally rely on inverse or inverse square root operations, which are implemented using the optimized \invSqrt method presented in Section~\ref{sec:construction}.
In this section, we detail the HE-based constructions of these operations and describe how our optimizations reduce multiplicative depth and improve runtime. For brevity, we omit the full description of skewness, as its computational structure closely mirrors that of kurtosis.

\subsection{Pre-normalization Scaling: Reducing Depth via Constant Folding}
\label{sec:prescaling}

To ensure stable bootstrapping and accurate polynomial approximation, inputs to \invSqrt must lie within a bounded domain—typically $[-1, 1]$ for bootstrapping and $[0, 2]$ for Chebyshev approximation. However, intermediate results such as variance and mean often exceed these ranges in practical applications. For example, the standard deviation in Z-score normalization can exceed $2$, placing its inverse square root outside the approximable range of Chebyshev polynomials.
To address this, we rescale intermediate values by multiplying the input ciphertexts with a constant factor $1/B$, where $B$ is a predefined bound. Unlike the conventional normalization method—which applies scaling after computing statistics such as variance and thus consumes an extra multiplicative level—our \textit{pre-normalization scaling} embeds the constant factor directly into the computation of mean and variance. This eliminates the additional multiplicative level overhead without compromising numerical correctness.
Since computing variance already requires two levels (due to squaring and averaging), our method keeps the total depth prior to applying \invSqrt at two instead of three. This transformation is shown in Equation~\ref{eq:2}.
{\small
    \begin{align}
\label{eq:2}
B \cdot \text{Var}(X) 
&= B \cdot \left( E[X^2] - E[X]^2 \right) \\
&= B \cdot \left ( \sum \left (X / \sqrt{N} \right)^2 - \left (\sum X / N \right)^2 \right ) \\
&= \sum \left( X / \sqrt{N/B} \right)^2 - \left( \sum X / (N/\sqrt{B}) \right)^2
\end{align}
}
This manipulation folds the constant $B$ into the normalization coefficients, allowing it to be applied earlier via plaintext multiplication. As a result, we reduce the cost of computing variance from three multiplicative levels to two before applying \invSqrt. To reflect this rescaling in the approximation, we modify the target function for inverse square root operation as:
\begin{equation}
\label{eq:3}
F(x, B) =
\begin{cases} 
\frac{1}{\sqrt{B} \cdot \sqrt{x+1}} & \text{if } x \geq -1 \\
0 & \text{otherwise}
\end{cases}
\end{equation}
We define \texttt{CryptoInvSqrt}$(x, B)$ as the Chebyshev approximation of $F(x, B)$ in Equation~\ref{eq:3}, ensuring numerical compatibility with the scaled variance. This strategy can also be applied to compute the square root of variance, as used in the coefficient of variation. For that, we define a scaled square root function:
\begin{equation}
\label{eq:44}
F(x, B) =
\begin{cases} 
\sqrt{B} \cdot \sqrt{x+1} & \text{if } x \geq -1 \\
0 & \text{otherwise}
\end{cases}
\end{equation}
Let \texttt{CryptoSqrt}$(x, B)$ denote the Chebyshev polynomial that approximates Equation~\ref{eq:44}.
We also apply this pre-normalization scaling technique to the mean operation:
{\small
\[
\text{Mean}(x, B) := \sum \left( \frac{x_i}{B \cdot N} \right) = \frac{1}{B} \cdot \text{Mean}(x)
\]
\[
\text{Variance}(x, B) := \sum \left( \left( \frac{x_i}{B \cdot\sqrt{N}} \right)^2 \right) - \left( \sum \left( \frac{x_i}{B \cdot N} \right) \right)^2 = \frac{1}{B^2} \cdot \text{Var}(x)
\]}

This constant-folding strategy reduces multiplicative depth by one level and is applicable to both mean and variance. An additional benefit of this method is that it enables the use of higher-degree Chebyshev polynomials under the same depth constraint. For example, in Z-score normalization, the maximum allowable Chebyshev degree is $2^8-1$ under depth 11 without pre-normalization scaling. With this technique, we can raise the polynomial degree to $2^9-1$ without increasing the number of bootstrapping rounds, thereby improving approximation accuracy without incurring additional computational cost.


\subsection{Z-Score Normalization (ZNorm)}
Z-score normalization is a standard statistical technique used to standardize values across datasets by removing the influence of differing means and variances. It is computed as the difference between a value and the mean, divided by the standard deviation:
{\small
\[
\text{ZNorm}(X) = \frac{X - \mu}{\sigma}
\]
}
Given that the standard deviation is the square root of the variance, computing its reciprocal reduces to evaluating the inverse square root of the variance. This can be efficiently approximated using \invSqrt, enabling low-depth Z-score normalization under HE. We apply a scaling factor $B$ during the variance operation to fit the input into the valid approximation range. The corresponding correction factor $B^2$ is passed to \texttt{CryptoInvSqrt} to ensure numerical consistency. 
The complete procedure is presented in Algorithm~\ref{alg:znorm}.
\begin{algorithm}[!ht]
\caption{HE-based Z-score Normalization (ZNorm)}
\label{alg:znorm}
\begin{algorithmic}[0]
\STATE \textbf{Input:} $ct_X$: Ciphertext of input vector $X$
\STATE \textbf{Output:} $ct_{znorm}$: Ciphertext of $\text{ZNorm}(X)$
\end{algorithmic}
\begin{algorithmic}[1]
\STATE $ct_{\mu} \leftarrow \texttt{Mean}(ct_X)$
\STATE $ct_{var} \leftarrow \texttt{Variance}(ct_X, B)$
\STATE $ct_{b} \leftarrow \texttt{CryptoInvSqrt}(ct_{var}, B^2)$
\STATE $ct_{a} \leftarrow \texttt{Sub}(C)(ct_X, ct_{\mu})$
\STATE $ct_{znorm} \leftarrow \texttt{Mul}(C)(ct_{a}, ct_{b})$
\STATE \textbf{return} $ct_{znorm}$
\end{algorithmic}
\end{algorithm}
\subsection{Kurtosis (Kurt)}
Kurtosis is a statistical measure that quantifies the heaviness of the tails of a distribution relative to a normal distribution.  It is defined as the ratio of the fourth central moment to the square of the variance:
{\small
\[
\text{Kurt}[X] = \frac{E[(x - E[X])^4]}{(E[(x - E[X])^2])^2}
\]
}
In HE, we compute the fourth central moment and the square of the variance homomorphically. The denominator is strictly non-negative, allowing us to apply an inverse square root approximation to its square for better depth efficiency.  
Algorithm~\ref{alg:kurtosis} outlines the full procedure.
\begin{algorithm}[!ht]
\caption{HE-based Kurtosis (Kurt)}
\label{alg:kurtosis}
\begin{algorithmic}[0]
\STATE \textbf{Input:} $ct_X$: Ciphertext of input vector $X$
\STATE \textbf{Output:} $ct_{kurt}$: Ciphertext of $\text{Kurt}(X)$
\end{algorithmic}

\begin{algorithmic}[1]
\STATE $ct_{\mu} \leftarrow \texttt{Mean}(ct_X)$
\STATE $ct_{x1} \leftarrow \texttt{Sub}(ct_X, ct_{\mu})$
\STATE $ct_{x2} \leftarrow \texttt{Mul}(C)(ct_{x1}, ct_{x1})$
\STATE $ct_{x4} \leftarrow \texttt{Mul}(C)(ct_{x2}, ct_{x2})$
\STATE $ct_{numerator} \leftarrow \texttt{Mean}(ct_{x4})$
\STATE $ct_{var} \leftarrow \texttt{Variance}(ct_X, B)$
\STATE $ct_{inv} \leftarrow \texttt{\invSqrt}(ct_{var}, B^2)$
\STATE $ct_{inv2} \leftarrow \texttt{Mul}(C)(ct_{inv}, ct_{inv})$
\STATE $ct_{inv4} \leftarrow \texttt{Mul}(C)(ct_{inv2}, ct_{inv2})$
\STATE $ct_{kurt} \leftarrow \texttt{Mul}(C)(ct_{numerator}, ct_{inv4})$
\STATE \textbf{return} $ct_{kurt}$
\end{algorithmic}
\end{algorithm}

\subsection{Coefficient of Variation (CV)}
The coefficient of variation (CV) is a statistical measure defined as the ratio of the standard deviation to the mean of a given dataset. Unlike variance or standard deviation, the denominator in CV is the mean, which may take negative values. This poses a challenge for polynomial approximations such as \texttt{\invSqrt}, which are typically defined only over non-negative domains. 

To address this, we extract the sign of the mean using a sign function, multiply it with the mean to ensure non-negativity before applying the inverse square root, and reapply the sign after computing the square of the inverse square root of the adjusted mean.

{\small
\[
\text{CV}(x) = \text{Std}(x) \cdot \frac{\text{sign}(\mu)}{|\mu|} 
= \text{Std}(x) \cdot \left(\invSqrt(\mu \cdot \text{sign}(\mu))\right)^{2} \cdot \text{sign}(\mu)
\]
}
Algorithm~\ref{alg:coeff_var} summarizes the implementation.

\begin{algorithm}[!ht]
\caption{HE-based Coefficient of Variation (CV)}
\label{alg:coeff_var}
\begin{algorithmic}[0]
\STATE \textbf{Input:} $ct_X$: Ciphertext of input vector $X$
\STATE \textbf{Output:} $ct_{cv}$: Ciphertext of CV(X)
\end{algorithmic}

\begin{algorithmic}[1]
\STATE $ct_{\mu} \leftarrow \texttt{Mean}(ct_X, B)$
\STATE $ct_{sign} \leftarrow \texttt{Sign}(ct_{\mu})$
\STATE $ct_{\mu\_pos} \leftarrow \texttt{Mul}(C)(ct_{\mu}, ct_{sign})$
\STATE $ct_{\mu\_invSqrt} \leftarrow \texttt{\invSqrt}(ct_{\mu\_pos}, B)$
\STATE $ct_{\mu\_inv} \leftarrow \texttt{Mul}(C)(ct_{\mu\_invSqrt}, ct_{\mu\_invSqrt})$
\STATE $ct_{var} \leftarrow \texttt{Variance}(ct_X, B)$
\STATE $ct_{std} \leftarrow \texttt{CryptoSqrt}(ct_{var}, B)$ \hfill 
\STATE $ct_{tmp} \leftarrow \texttt{Mul}(C)(ct_{std}, ct_{\mu\_inv})$
\STATE $ct_{cv} \leftarrow \texttt{Mul}(C)(ct_{tmp}, ct_{sign})$
\STATE \textbf{return} $ct_{cv}$
\end{algorithmic}
\end{algorithm}

\subsection{Pearson Correlation Coefficient (PCC)}
The Pearson correlation coefficient (PCC) is a statistical measure of linear dependence between two random variables.  
It is defined as the ratio of their covariance to the product of their standard deviations:
{\small
\[
\text{PCC}(X, Y) = \frac{\text{Cov}(X, Y)}{\sqrt{\text{Var}(X)} \cdot \sqrt{\text{Var}(Y)}}
\]
}
Lee et al.~\cite{lee2023heaan} proposed a homomorphic algorithm for computing PCC using the CKKS scheme. In our implementation, we compute the centered versions of $X$ and $Y$, then apply mean, multiplication, and inverse square root operations homomorphically. The inverse square roots are approximated using \invSqrt, as in other statistical functions. The overall procedure follows Algorithm~3 in \cite{lee2023heaan} for detailed pseudocode.

\section{Experiment}
\label{sec:experiment}

\subsection{Experimental Setup}
\label{sec:experiment_setup}
We evaluate the precision and computational efficiency of \sysname under standard CKKS homomorphic encryption settings.  All experiments are implemented using Lattigo v6~\cite{lattigo}, an open-source HE library written in Go that supports approximate arithmetic over real numbers.  For a fair and consistent comparison, we reimplemented the inverse $n$-th square root algorithms proposed in \panda and \hstat within the same Lattigo environment. \\
\textbf{Hardware Configuration.} 
All experiments were conducted on a server equipped with an Intel(R) Xeon(R) Gold 6248R CPU @ 3.00GHz processor, 64GB RAM, and 500GB SSD. All implementations were executed in single-thread mode to ensure consistency across all evaluations. \\
\textbf{Parameter Settings.} 
We set the ring dimension (polynomial degree) to $N = 2^{16}$ with a maximum depth of 27 and the scale factor to $\Delta = 2^{40}$. The total modulus size $\log_2(PQ)$ is set to 1443 bits, which supports up to 11 levels of multiplicative depth before requiring bootstrapping. For bootstrapping, $\log_2(PQ)$ is set to 744 bits. The number of slots is $32,768$, half of $N$. This configuration satisfies the 128-bit security level under the standard RLWE hardness assumptions~\cite{bossuat2024security}.

\subsection{Experiment 1: Performance Comparison of Inverse Square Root Operation}
\begin{figure}[!h]
\centerline{\includegraphics[width=0.8\columnwidth]{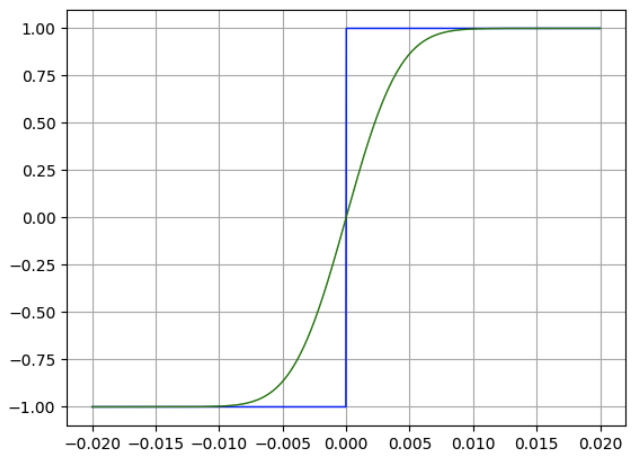}}
\caption{Comparison of step functions: $G_3^{(7)}(x)$ (green) vs. minimax approximation (blue).}
\label{fig:step_functions}
\end{figure}

The inverse square root operation is a critical component in computing fundamental statistical measures such as Z-score normalization, kurtosis, skewness, and the Pearson correlation coefficient in \sysname. Given its central role, we evaluate the accuracy and efficiency of the proposed \invSqrt method in comparison to existing techniques, specifically those proposed in \hstat and \panda. Qu and Xu~\cite{qu2023improvements} proposed rational and Taylor-based initializations of $1/\sqrt{x}$. However, as their polynomial coefficients are not disclosed, a direct numerical reproduction is infeasible, and we therefore restrict our comparison to qualitative discussion.
In this experiment, we measure the MRE of inverse square root operations across three methods: \hstat, \panda, and our proposed \invSqrt.
Panda et al.~\cite{panda2022polynomial} utilize a function $G_3^{(7)}(x)$ as a \textit{sign} function to generate a suitable initial value for Newton’s method.  
$G_3^{(7)}(x)$ denotes the 7-fold composition of the degree-7 polynomial $g_3(x)$:
{\small
\[
g_3(x) = \frac{35}{16}x - \frac{35}{16}x^3 + \frac{21}{16}x^5 - \frac{5}{16}x^7.
\]
}
However, $G_3^{(7)}(x)$ exhibits instability in the narrow interval $|x| \leq 0.01$, leading to poor approximation behavior near zero. To address this, we adopt a minimax-optimized step function~\cite{lee2022optimization}, which, while requiring 5 additional multiplicative levels, achieves significantly higher precision over the entire input domain.
Figure~\ref{fig:step_functions} compares $G_3^{(7)}(x)$ (green) with the minimax approximation (blue).

Both \panda and \invSqrt are designed to operate directly over the input domain $[0.001, 100.0]$ without any preconditioning.  
In contrast, the original \hstat method requires inputs in the interval $(0, 1]$. To enable a fair comparison, we scale the inputs for \hstat by a factor of $1/100$, and subsequently rescale the outputs by $\sqrt{100}$ to align with the full domain of the inverse square root function.
To assess approximation accuracy across a wide dynamic range, we evaluate the MRE over 32,768 real-valued inputs sampled from $[0.001, 100]$. The domain is partitioned into two subranges to capture both low- and high-value behaviors: 16,384 inputs in $[0.001, 1.0]$ (spacing $\approx 6.10 \times 10^{-5}$), and 16,384 in $[1.0, 100]$ (spacing $\approx 6.04 \times 10^{-3}$).

Table~\ref{table:invsqrt_performance_table} summarizes the evaluation results. Here, $\#\text{BTS}$ denotes the number of bootstrapping operations, and Runtime (s) is measured using single-core execution. All values are averaged over 10 runs, with standard deviations shown in parentheses. Unlike \panda, which supports only the narrow input domain $[0, 1]$, \hstat requires normalization when operating over broader ranges. To ensure compatibility, we scale the input data by a constant factor of $1/B$, where $B = 100$. This same factor is also applied inside \invSqrt to maintain consistency with the transformed domain.

\begin{table}[!ht]
\caption{Performance comparison of inverse square root operation over the input domain $[0.001, 100]$. The parameter \textit{B} denotes the constant scaling factor. All values are averaged over ten runs; values in parentheses represent standard deviations.}
\centering\resizebox{1\linewidth}{!}{
\renewcommand{\arraystretch}{1.3}
\begin{tabular}{c|cccc}
\toprule[1.3 pt]
\textbf{Method} & $B$& $\#\text{BTS}$ & \textbf{MRE} & \textbf{Runtime (s)} \\
\hline
\hstat & - & 5 & $5.29 \times 10^{-3}$ ( $8.53 \times 10^{-8}$ ) & 225.02 (3.18) \\
\panda & 100.0 & 6 & $3.73 \times 10^{-4}$ ( $1.68 \times 10^{-9}$ ) & 273.48 (5.29) \\
\textbf{\invSqrt} & 100.0 & 2 & $5.08 \times 10^{-5}$ ( $1.75 \times 10^{-9}$ ) & 94.73 (1.05) \\
\toprule[1.3 pt]
\end{tabular}}
\label{table:invsqrt_performance_table}
\end{table}
As shown in Table~\ref{table:invsqrt_performance_table}, \invSqrt achieves the highest precision among all methods, with a MRE of $5.08 \times 10^{-5}$. While all methods adopt Newton’s method for inverse square root approximation, \invSqrt utilizes a Chebyshev-based strategy to determine an optimized initial guess, significantly reducing the number of Newton iterations and associated bootstrapping operations.
As shown in Table~\ref{table:op_comparision}, bootstrapping is approximately $221.44\times$ slower than homomorphic multiplication. Consequently, \invSqrt requires fewer bootstrapping rounds and achieves the fastest runtime. To accommodate the wide input domain $[0.001, 100]$, we set the constant scaling factor $B$ to 100, enabling accurate approximation within the Chebyshev-valid range.

Compared to \panda, \invSqrt achieves approximately $7.34\times$ lower MRE while consuming less than half the number of bootstrapping operations. Relative to \hstat, \invSqrt delivers over $104.13\times$ better accuracy and reduces runtime by a factor of $2.38\times$ (94.73s vs. 225.02s). These results highlight the effectiveness of our Chebyshev-based initialization in reducing both multiplicative depth and latency in homomorphic inverse square root operation.

Compared to Qu and Xu~\cite{qu2023improvements}, whose methods required depths of at least 35–36 (three or more bootstraps in our CKKS setting), our Chebyshev-based initialization completes with only two bootstraps. Although the ranges differ, this qualitative comparison indicates lower bootstrapping overhead than prior strategies.

\subsection{Experiment 2: Accuracy on Large-Scale Dataset}
In this experiment, we evaluate the MRE of five core statistical operations in \sysname: Z-score normalization, skewness, kurtosis, PCC, and CV. For scalar-valued metrics (e.g., kurtosis, skewness), MRE corresponds to the relative error between the decrypted result and its plaintext reference. Z-score normalization is applied to raw input distributions with wide dynamic ranges, while the remaining operations are computed over normalized data. Since Chebyshev polynomial approximation suffers from increased error over wide domains, we evaluate Z-score normalization over $[0, 100]$ and the other measures over $[0, 20]$. Each experiment is conducted on one million independently sampled real-valued inputs to simulate large-scale data processing. Table~\ref{table:stat_result} summarizes the accuracy and runtime.

\begin{table}[ht]
\small
\centering
\caption{Accuracy and efficiency of Z-score normalization evaluated over the domain $[0, 100]$, and the remaining four statistical measures over $[0, 20]$. All values are averaged over 10 trials; values in parentheses denote standard deviations. The parameter $B$ denotes the constant scaling factor used for normalization.}
\begin{tabular}{lcccc}
\toprule
\textbf{Measure} & B & \#\text{BTS}& \textbf{MRE} & \textbf{Runtime (s)} \\
\midrule
ZNorm & 100 & 2 & $4.18 \times 10^{-5}$ ($6.06 \times 10^{-6}$) & 141.29 (1.55) \\
Skew & 20 & 2 &  $8.12 \times 10^{-3}$ ($1.41 \times 10^{-2}$) & 154.10 (1.61) \\
Kurt & 20 & 2 & $3.73 \times 10^{-4}$ ($6.97 \times 10^{-6}$) & 154.70 (1.34) \\
CV & 20 & 7 & $1.25 \times 10^{-4}$ ($9.70 \times 10^{-5}$) & 311.08 (2.94) \\
PCC & 20 & 4 & $2.62 \times 10^{-4}$ ($3.21 \times 10^{-5}$) & 289.86 (3.47) \\
\bottomrule
\end{tabular}
\label{table:stat_result}
\vspace{1mm}
\footnotesize{\textit{Abbreviations:} ZNorm = Z-score normalization, skew = skewness, kurt = kurtosis}
\end{table}

Table~\ref{table:stat_result} shows that Z-score normalization achieves high accuracy, with an MRE of $4.18 \times 10^{-5}$ over the wide input domain $[0, 100]$. To ensure compatibility with this domain, we apply a constant scaling factor of $B = 100$ to align the inputs to the Chebyshev-valid range.

Among the five operations, four—Z-score normalization, skewness, kurtosis, and PCC—require computing the inverse square root of variance. In contrast, the CV additionally involves computing the inverse of the mean. These intermediate values often exceed the valid Chebyshev approximation domain $[0, 2]$, which may lead to substantial approximation error. To address this, we normalize the variance and mean using a constant scaling factor $B = 20$, applying $1/B^2 = 1/400$ and $1/B = 1/20$, respectively. This transformation ensures compatibility with the approximation domain, while correctness is preserved through the multiplicative homomorphism of CKKS and symmetric post-processing.

Unlike variance, which is always non-negative, the mean can be negative. To correctly compute the inverse in such cases, we adopt a sign-aware approach: we extract the sign using the minimax approximation~\cite{lee2021minimax}, apply \invSqrt to the absolute value, and reapply the sign. This is implemented using a minimax-optimized polynomial of multiplicative depth 32, requiring three bootstrapping rounds. As a result, CV incurs the highest bootstrapping cost (seven rounds) due to its multi-phase computation, including standard deviation calculation, sign-aware inverse mean approximation, and additional polynomial evaluations. In contrast, the other measures require at most four bootstrapping rounds, reflecting their lower depth complexity and higher runtime efficiency.

\subsection{Experiment 3: Evaluation on Real-world Datasets}

To evaluate the practical applicability of \sysname in real-world settings, we conduct experiments on two widely used datasets: the UCI Adult Income Dataset~\cite{adult_2} and the Medical Cost Insurance Dataset~\cite{akter2018investigation}.  
For brevity, we refer to them as \textit{Adult} and \textit{Insurance}, respectively.

\subsubsection{Dataset Descriptions}
The \textit{Adult} dataset contains 48,842 records with 14 attributes, including \texttt{age}, \texttt{hours-per-week}, and \texttt{education-} \texttt{num}. It is commonly used to predict whether an individual's income exceeds \$50,000.  
The \textit{Insurance} dataset consists of 1,338 samples across 7 features such as \texttt{age}, \texttt{bmi}, \texttt{smoker}, and \texttt{charges}, and is often used in regression and cost modeling tasks.

Using \sysname, we compute several statistical measures—including Z-score normalization, skewness, kurtosis, CV, and PCC—on selected features from both datasets.  
Decrypted outputs are compared to plaintext baselines to assess numerical accuracy.

\subsubsection{Evaluation on the UCI Adult Income Dataset}
We select three continuous-valued features from the Adult income dataset (\textit{Adult} dataset): \texttt{age}, \texttt{education-num}, and \texttt{hours-per-week}. For each feature, we compute the mean relative error (MRE) for Z-score normalization as well as for four additional statistical measures: skewness, kurtosis (reported as excess kurtosis), CV, and PCC. 
In addition, we compute the PCC between the feature pairs \texttt{(age, hours-per-week)} and \texttt{(age, education-num)}. We empirically set the constant scaling factor $B$ to 50, based on the observation that the means of the selected features—particularly \texttt{hours-per-week} typically fall between 30 and 50. The results are summarized in Table~\ref{table:adult_evaluation}.
\begin{table}[ht]
\centering
\caption{Evaluation of statistical operations over the \textit{Adult} dataset. We report MRE for Z-score normalization (ZNorm), kurtosis (Kurt), skewness (Skew), and coefficient of variation (CV) across selected features. PCC denotes the Pearson correlation coefficient between feature pairs. The parameter $B$ denotes the constant scaling factor. All values are averaged over ten trials; values in parentheses indicate standard deviations.}
\label{table:adult_evaluation}
\resizebox{1\linewidth}{!}{
\begin{tabular}{l|lcccc}
\toprule
\textbf{Measure} & \textbf{Feature(s)} &$B$ & \textbf{Output} & \textbf{MRE} & \textbf{Runtime (s)} \\
\midrule
ZNorm & AGE & 50 & - & $2.47 \times 10^{-5}$ ( $3.39 \times 10^{-21}$ ) & 110.18 (2.42) \\
      & EDU & 50 & - & $1.02 \times 10^{-4}$ ( $1.36 \times 10^{-20}$ ) & 110.03 (1.40) \\
      & HPW & 50 & - & $7.62 \times 10^{-5}$ ( $1.36 \times 10^{-20}$ ) & 109.73 (1.71) \\
\midrule
Skew  & AGE & 50 & 0.5576 & $7.92 \times 10^{-5}$ ( $1.36 \times 10^{-20}$ ) & 113.92 (1.83) \\
      & EDU & 50 & -0.3165 & $2.50 \times 10^{-4}$ (0.00)                      & 112.32 (2.02) \\
      & HPW & 50 & 0.2387 & $1.38 \times 10^{-4}$ (0.00)                      & 111.81 (1.68) \\
\midrule
Kurt  & AGE  & 50 & -0.1844 & $4.81 \times 10^{-3}$ (0.00) & 113.20 (2.07) \\
      & EDU  & 50 & 0.6256 & $2.14 \times 10^{-3}$ (0.00) & 113.02 (1.58) \\
      & HPW  & 50 & 2.9506 & $7.76 \times 10^{-4}$ (0.00) & 112.36 (1.52) \\
\midrule
CV    & AGE  & 50 & 0.3548 & $4.39 \times 10^{-5}$ ( $6.78 \times 10^{-21}$ ) & 297.60 (3.84) \\
      & EDU  & 50 & 0.2551 & $1.10 \times 10^{-3}$ ( $2.17 \times 10^{-19}$ ) & 295.72 (3.92) \\
      & HPW  & 50 & 0.3065 & $4.82 \times 10^{-5}$ ( $6.78 \times 10^{-21}$ ) & 295.68 (3.08) \\
\midrule
PCC   & AGE vs HPW & 50  & 0.0716 & $1.40 \times 10^{-4}$ ( $2.71 \times 10^{-20}$ ) & 222.39 (3.05) \\
      & AGE vs EDU & 50 & 0.0309 & $1.01 \times 10^{-4}$ (0.00)    & 222.38 (3.35) \\
\bottomrule
\end{tabular}
}
\vspace{1mm}
\footnotesize{\textit{Abbreviations:} AGE = age, EDU = education-num, HPW = hours-per-week}
\end{table}
As shown in Table~\ref{table:adult_evaluation}, \sysname maintains high accuracy across all metrics. Z-score normalization, skewness, and kurtosis yield MREs below $3 \times 10^{-3}$, while PCC results show errors consistently under $1.4 \times 10^{-4}$. CV results vary depending on feature scale, with increased error observed in \texttt{education-num}, which has a relatively small mean. This behavior aligns with the known sensitivity of CV to small denominators. 

\subsubsection{Evaluation on the Medical Cost Insurance Dataset}
\begin{figure}[htbp]
\centerline{\includegraphics[width=0.8\columnwidth]{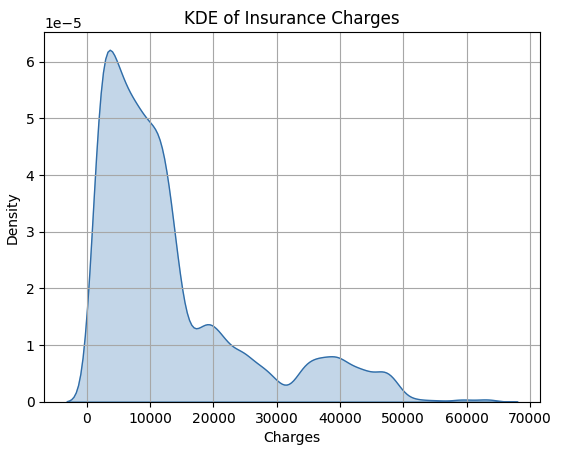}}
\caption{Kernel density estimation (KDE) of \textit{charges} in the \textit{Insurance} dataset.}
\label{fig:kde_charges}
\end{figure}

To assess the applicability of \sysname to real-world privacy-preserving statistical analysis tasks, we conduct experiments on the Medical Cost Insurance Dataset (\textit{Insurance} dataset), which includes sensitive personal and medical information often used in actuarial studies. We select three representative features: \texttt{age}, \texttt{bmi}, and \texttt{smoker}, which are frequently analyzed in medical cost prediction. Categorical or discrete features such as \texttt{sex}, \texttt{children}, and \texttt{region} are excluded to simplify encrypted operation and ensure reproducibility.
Both \texttt{age} and \texttt{bmi} are continuous variables ranging from 18 to 64 and 15.96 to 53.13, respectively.  The \texttt{smoker} attribute is binary, and we map it to numerical values (yes $\rightarrow$ 1, no $\rightarrow$ 0) to enable encrypted computation of PCC.  The target variable \texttt{charges} ranges from 1121.87 to 63770.43; we scale it by a factor of $1/1000$ before encryption to match the CKKS dynamic range.  Since the PCC is scale-invariant, this transformation does not affect the final result. This configuration enables privacy-preserving statistical analysis on encrypted features and target variables, allowing us to extract meaningful patterns without exposing sensitive data. We set $B = 100$ for Z-score normalization, and $B = 20$ for all other evaluations. Table~\ref{table:insurance_evaluation} summarizes the evaluation results.
\begin{table}[ht]
\centering
\caption{Evaluation of statistical metrics using \sysname over the \textit{Insurance} dataset.  
We report the mean relative error (MRE) for each statistical function compared to the plaintext result. PCC denotes the PCC between the target feature (\texttt{charges}) and selected predictors. The parameter $B$ denotes the constant scaling factor. Kurtosis is reported as excess kurtosis (i.e., normal kurtosis minus 3).}
\label{table:insurance_evaluation}
\resizebox{1\linewidth}{!}{
\begin{tabular}{l|lcccc}
\toprule
\textbf{Measure} & \textbf{Feature(s)} & $B$ & \textbf{Output} & \textbf{MRE} & \textbf{Runtime (s)} \\
\midrule
ZNorm & charges & 100 & -                  & $3.81 \times 10^{-5}$ ( $0.00$ )            & 108.26 (3.34) \\
\midrule
Skew   & charges & 20 & 1.5143                  & $8.67 \times 10^{-5}$ ( $1.36 \times 10^{-20}$ ) & 105.52 (2.36) \\
Kurt   & charges & 20& 1.5966 & $6.08 \times 10^{-4}$ ( $1.08 \times 10^{-19}$ ) & 104.94 (2.45) \\
CV     & charges & 20& 0.9123 & $5.54 \times 10^{-5}$ ( $6.78 \times 10^{-21}$ ) & 279.02 (5.11) \\
\midrule
PCC    & AGE vs charges & 20& 0.2990  & $2.22 \times 10^{-4}$ ( $0.00$ ) & 207.87 (4.73) \\
       & BMI vs charges  & 20& 0.1983  & $7.33 \times 10^{-5}$ ( $0.00$ ) & 207.85 (3.64) \\
       & SMOKER vs charges & 20& 0.7873 & $2.86 \times 10^{-4}$ ( $5.42 \times 10^{-20}$ ) & 209.98 (2.07) \\
\bottomrule
\end{tabular}
}
\end{table}
As shown in Table~\ref{table:insurance_evaluation}, \sysname achieves high accuracy across all evaluated statistical functions. Z-score normalization, skewness, and CV produce MREs below $8 \times 10^{-4}$, highlighting the framework’s robustness even for high-variance numerical attributes. Pearson correlation values involving \texttt{age} and \texttt{bmi} also maintain low error.  In contrast, the PCC involving the binary \texttt{smoker} feature shows higher error due to its limited dynamic range, which affects the stability of the denominator in correlation operation. As shown in Table~\ref{table:insurance_evaluation}, all relative errors lower than $3 \times 10^{-4}$, demonstrating the high numerical precision of \sysname in encrypted computation.
The skewness of \textit{charges} is measured as 1.5143, indicating a right-skewed distribution. This suggests that most insurance fees are concentrated near the lower range, while a small subset of individuals are charged substantially higher fees. The excess kurtosis is 1.5966, implying a leptokurtic distribution with heavy tails. This reflects the presence of extreme outliers—individuals who incur unusually high charges, a common pattern in real-world insurance data.
The CV is 0.9123, suggesting that the standard deviation is nearly as large as the mean. This highlights the substantial variability in insurance fees across the population, and underscores the need for secure statistical tools in sensitive domains. Z-score normalization achieves MRE well below $10^{-4}$, validating \sysname’s ability to support encrypted feature scaling with high precision.  
Figure~\ref{fig:kde_charges} illustrates the kernel density estimation (KDE) of \textit{charges}, visually confirming the right-skewness and heavy-tailed structure observed in our numerical evaluation.

In the PCC evaluation, the correlation between \texttt{smoker} and \texttt{charges} is 0.7873, indicating a strong positive association. This indicates that smoking status is a dominant factor affecting medical costs. The PCC between \texttt{age} and \texttt{charges} is 0.2990, reflecting a moderate positive correlation—insurance fees tend to increase with age.  
Meanwhile, the PCC for \texttt{bmi} is 0.1983, suggesting a weaker but still positive relationship. These results demonstrate that \sysname enables accurate and meaningful higher-order statistical analysis on encrypted medical cost data. The ability to perform normalization, distributional analysis, and correlation estimation directly in the encrypted domain underscores the practical value of \sysname in privacy-preserving healthcare analytics.

\section{Discussion and Future Work}
\label{sec:discussion_futurework}
In this work, we focus on the Chebyshev approximation for determining suitable initial values in Newton's iteration. Beyond engineering optimizations, \sysname{} introduces key technical innovations, including pre-normalization scaling to reduce multiplicative depth, Chebyshev-based initialization for fast \invSqrt convergence.
However, potential vulnerabilities arising from polynomial approximation errors—such as information leakage through output deviation patterns—were not deeply analyzed in this study. In future work, we plan to conduct a comprehensive security analysis of such approximation-based vulnerabilities.
Currently, \sysname{} supports only five statistical operations. We plan to investigate more optimized operations, such as minimum and maximum. The computation of CV incurs a considerably large runtime due to the use of the sign function, which requires three bootstrapping operations; optimizing the CV algorithm to reduce this cost is therefore necessary.
While our current focus is on encrypted statistical analysis, extending \sysname{} to privacy-preserving machine learning (ML) pipelines is a natural next step. This includes enabling secure feature selection, model training, and inference on encrypted datasets.
At present, \sysname{} is implemented using Lattigo, a Go-based HE library that operates in CPU-only environments. Since HE operations are computationally intensive with limited throughput in such settings, we plan to explore hardware acceleration via GPU-based HE libraries and dedicated computing platforms such as FPGAs or ASICs.
Furthermore, we aim to extend \sysname{} to support secure multi-party computation (MPC), federated analytics, and differential privacy (DP) mechanisms. Such integration will broaden the applicability of \sysname{} in collaborative, distributed, and privacy-regulated environments.

\section{Conclusion}
\label{sec:conclusion}
In this work, we presented \sysname{}, a homomorphic encryption (HE)-based framework for privacy-preserving statistical analysis. \sysname{} supports five fundamental statistical measures—Z-score normalization, skewness, kurtosis, coefficient of variation, and Pearson correlation coefficient—directly over encrypted data. 
A key contribution is our optimized inverse square root operation, achieving $7.34\times$ lower mean relative error (MRE) and $2.38\times$ faster runtime compared to prior approaches such as \panda{} and \hstat{}. This reduces multiplicative depth and bootstrapping frequency, enabling practical computation of higher-order statistics under HE. We also proposed a \textit{pre-normalization scaling} technique that eliminates the additional multiplicative level without additional cost. 
We validated \sysname{} on two real-world datasets—\textit{Adult} and \textit{Insurance}—and demonstrated accurate and efficient encrypted computation across diverse statistical tasks. These results confirm that \sysname{} provides a practical and scalable solution for secure data analysis in privacy-sensitive domains.

\section*{GenAI Usage Disclosure}
This paper was entirely written by the authors. During the preparation of the manuscript, we used ChatGPT (developed by OpenAI) solely for the purpose of grammar correction, word choice refinement, and improvement of sentence clarity. No content, sentence, or paragraph was generated by the model without substantial author input. All scientific claims, experimental results, and explanations were written and validated by the authors.
\bibliography{acmart} 

\end{document}